# RAFFAELLO CAVERNI (1837 - 1900) AND THE SOCIETY FOR THE PROGRESS OF THE SCIENCES: AN INDEPENDENT PRIEST CRITICIZED BY THE LAY SCIENTISTS*

Dino Boccaletti

(Former professor of Celestial Mechanics - Sapienza University of Rome)

ABSTRACT

Raffaello Caverni, a Catholic priest, was a truly lay and anti-establishment intellectual in his opinions both on Darwin and on Galileo. He opposed the mythicization of Galileo, as a rule in Italy after the unification, even though he considered Galileo a great scientist.
As a consequence the scientific community of that time, under the influence of Antonio Favaro, bitterly censured his work *Storia del Metodo Sperimentale in Italia*.In this way, Caverni's book was removed from the scientific debate in Italy for at least forty years.

INTRODUCTION

On 17 April 1917, the Division of History of Sciences of the X Conference of the Italian Society for the Progress of the Sciences, convened in Florence under the chairmanship of Enrico D'Ovidio with Aldo Mieli as secretary, approved the following agenda:
La Sezione di Storia delle Scienze, udita la comunicazione del Prof. Carlo Del Lungo sopra la *Storia del metodo sperimentale in Italia* di R. Caverni, di fronte a questa e ad altri rinnovati tentativi antigalileiani, conferma il voto già espresso in questo Congresso, che cioè si ristampino in nuova edizione nazionale le opere di Galileo, mettendole in vendita e diffondendole il più possibile in Italia e all'estero; fa inoltre voto che per iniziativa della Società Italiana per il Progresso delle Scienze sia fatta una recensione critica della storia del Caverni, nella quale vengano messi in chiaro gli intendimenti ed i mezzi adoperati dall'autore nel giudicare l'opera di Galileo, e che a tale pubblicazione, fatta possibilmente in più lingue, sia data amplissima diffusione fra tutti gli studiosi in Italia e all'estero.[1]
At this point the man in the street wonders who Raffaello Caverni was, to get such a treatment. Why a prestigious institution, which in the Division of History of Sciences had members as Federigo Enriques (incidentally, also present in that meeting), Antonio Favaro and Roberto Marcolongo, to limit ourselves to the better known academicians great experts of the work of Galileo, feels the need and takes the responsibility for approving such an agenda? It is quite usual, we dare say physiologic, that individual scholars criticize the work of another scholar. On the contrary, it is rather unusual that a prestigious Society starts a sort of trial against a work (of six volumes, five appeared between 1891 and 1898 and the sixth perhaps in 1917 but bearing the date 1900) of history of science.
The result of this operation was that Caverni's work was completely segregated in the Italian culture for at least forty years.
To arrive at an understanding of the origin of the *affaire* and to contextualize it in the cultural *milieu* of the post-unification of Italy, it is necessary to recall some elements which allow us to get an idea of the socio-cultural atmosphere of that time and, obviously, also to take a quick look at the biography of Raffaello Caverni. Let us begin from the first point.

---

* Communication held at the meeting *Dall'Astronomia alla Cosmologia – Domus Galilaeana – Pisa,* November 19th, 2010.

[1] The Division of History of Sciences, having heard the communication of prof. Carlo Del Lungo regarding the *History of the experimental method in Italy* of R. Caverni, facing this and other reiterated anti-Galilean attempts, confirms the vow already made in this Conference, i.e. Galileo's works must be reprinted in a new national edition, putting them up for sale and disseminating them as much as possible in Italy and abroad; and vows as well that on the initiative of the Italian Society for the Progress of the Sciences a critical review of Caverni's history will be edited, making clear the intentions and the means used by the author in judging the work of Galileo, and that this review will be published, if possible, in various languages and widely spread among all the scholars in Italy and abroad.

THE MYTH OF GALILEO IN THE ITALIAN CULTURE AFTER THE UNITY OF ITALY

First of all let us recall that on the morrow of the unification of Italy there was the problem of *making the Italians,* according to the well-known sentence of Massimo D'Azeglio. A part of this operation, if we stay within the cultural ambit, consisted in making a lot of room for commemorations, celebrations, etc. of *great Italians* (Colombo, Galileo, to mention the greatest ones). Inevitably, this led to a certain mythicization of those personages. In the case of Galileo the construction of the myth turned out rather easy.
Galileo ebbe infatti le qualità naturali di un leader, compreso un carattere autoritario non privo di arroganza; come tale riuscì a creare una vera e propria scuola legata eminentemente al prestigio e all'autorità della sua persona e non tanto a una dottrina precisa, dato che il suo pensiero non si espresse mai in forma compiuta e sistematica (Micheli)[2]. The trial and the consequent condemnation, among other things, rendered Galileo a hero of the free thought and the new Italy could make use of his image as an anti-Vatican symbol. Indeed, the celebrations held in Pisa (1864) on the occasion of the third centennial anniversary of Galileo's birth primed polemics between Liberals and Clericals with journalistic after-effects. In the Grand Ducky of Tuscany, since Viviani's *Vita di Galileo* (1654) onwards, the memory of Galileo handed on by his pupils and continuers had ever been kept alive. To this, it should be added the action of the Tuscan scholars of the XVIII century (Giovanni Targioni Tozzetti and Gio. Battista Clementi Nelli) and of the exile Guglielmo Libri who in his work *Histoire des sciences mathématiques en Italie* (1838) brought Galileo into great relief also for his contribution to the establishment of mathematics as a doctrine absolutely necessary to the progress of science. Moreover, between 1842 and 1856, a new edition of Galileo's works in 15 volumes, at that time presented as the first complete edition and sponsored by the grand duke Leopoldo II, was published in Florence edited by Eugenio Albèri. In the first volume of the Supplement of this edition, Albèri already dedicated about fifty pages to an examination of the *Opinions and Judgments of F. Arago about G. Galilei* ( Arago was author of new and odd censures - as Albèri wrote - in a biography of Galileo inserted in the III tome of the posthumous complete edition of his works). Therefore, we can take this date (1856) as the beginning of the struggle of the Italian scholars in the grand-ducal Tuscany against those who will be called the *detractors of Galileo* (besides, an epithet borrowed from the very words of Galileo in the incipit of the *Assayer* written as a letter to the Lynceus academician don Virginio Cesarini).

---

[2] In fact, Galileo had the natural qualities of a leader, including an authoritative character not devoid of arroganza; in this way, he succeeded in creating a real school eminently tied to his personal prestige and authority and not to precise doctrine, since his thought was never expressed in a complete and systematic form (Micheli).

We can say that the fight against the detractors, at least within the limits we shall discuss later on, lasted about a century.

Obviously, we are speaking here about the scholars of Galileo, since on a popular level the mythical image is destined to survive. Before discussing the question which involves Raffaello Caverni in the group of the so-called detractors of Galileo, nay the greatest detractor of Galileo (according to Favaro), let us explore, as much as possible, his biography.

RAFFAELLO CAVERNI

The biography of Raffaello Caverni can be exhausted in few lines if we limit ourselves to report, say, the personal events.
He was born in a small Tuscan village (S. Quirico di Montelupo) on 12 March 1837. When a boy, already directed to the ecclesiastical ministry, he attended the *Scuole Pie* of Florence and then the *Istituto Ximeniano* under excellent teachers of Sciences. Even before entering the Church, he was sent to teach (during the school year 1859-60) in the seminary of Firenzuola. He entered the church on 2 June 1860. He has taught in Firenzuola for 10 years, during which, besides to teach, he deepened his studies of physics and natural sciences. At the end of 1870 he was sent as a parish priest in the parish of Quarata (Quarata Antellese) in Val d'Ema, municipality of Bagno a Ripoli, close to Florence.
He remained in this parish the following thirty years and there died on 30 January 1900. Fortunately for him, the parish, comparatively small, gave him much time free. He dedicated this time to studies and to frequenting the National Library of Florence. Thus, the thirty years spent at Quarata were the years of his production as historian and writer of popular science. He began his popularizing activity through notes with historical parts (Scientific Recreations) on Florentine magazines of the time. Later on, some of these writings were collected in volume. The first book among those dedicated to popularize natural sciences was *Problemi naturali di Galileo Galilei e altri autori della sua scuola*[3], published by G. C. Sansoni in 1874.
Other books followed, among which also *Dizionarietto di voci e di modi nella Divina Commedia dell'uso popolare toscano* [4] (1878), which we mention in order to emphasize the large variety of interests of Caverni. In 1875 and 1876 Caverni published on the «Rivista Universale», a catholic cultural magazine also dedicated to topics of scientific and philosophical interest of the day, articles under the title *On the philosophy of Natural Sciences*. Collected in a volume, these articles were published in 1877 with the title *De' nuovi studi della filosofia, discorsi di R. C. ad un giovane studente*[5]. «Civiltà Cattolica» (the influential Jesuits review) concentrated doggedly on this book, and, since 1878, dedicated a long series of articles to the Darwinism (signed by father Pietro Caterini S.J.), which was just the subject dealt with in Caverni's book.
Before telling the consequences of this, let us briefly try to make clear the context in which the question was included.
As it is known, the work of Darwin *The origin of Species by means of natural selection* was published in English at the end of 1859 and translated in Italian by Giovanni Canestrini and published by *Zanichelli* - Modena on 1864.

---

[3] *Natural Problems of Galileo Galilei and other authors of his school*
[4] *Little Dictionary of voices and moods of the Divine Comedy in the Tuscan vernacular*
[5] *On the new studies of the philosophy, talks of R. C. to a young student.*

In Italy, a heated debate arose on that work and, particularly at the beginning, the Darwinism was received in the catholic *milieu* with deep hostility and contempt. The Florentine Catholic *intelligentsia,* beginning from the Piarist Giovanni Antonelli (1818-1872), was particularly hard on Darwinian theories and also some renowned intellectuals of the time, as Gino Capponi, Raffaello Lambruschini, Terenzio Mamiani, adopted the same position.

On the contrary, the priest Raffaello Caverni, regarding the Darwinian theories and their possible clash with the letter of the Bible, applies a method which, we dare say, faithfully follows the one carried on by Galileo in the famous *Letter to Madam Cristina of Lorena* (1615). In fact he says ... imbattutomi in quest'acre disputazione, che tien gli uomini da qualche anno agitati intorno all'origine delle specie animali, volli più riposatamente fermarmi a esaminare queste nuove dottrine, propugnate con tanto ardore da una falange di naturalisti, duce della quale è un nome celebre, Carlo Darwin. E riscontrando le dottrine di lui con la Genesi [.... ] mi pare aver trovato che tutt'altro ch'essere le nuove dottrine del naturalista inglese opposte a quel che leggesi nel Libro di Mosè, vi s'accomodino invece con molto miglior ordine che nelle interpretazioni degli antichi esegeti [6]. This point was not swallowed by the hierarchies. On denunciation of the Archbishop of Florence, Eugenio Cecconi, the Congregation of the Index began to move on November 1877. On May 1878, the eminent Dominican Tommaso Maria Zigliara presented a report of 99 pages on Caverni's book. On July 1878, the Congregation of the Index decided to insert Caverni's book in the *Index Librorum Prohibitorum* (decree of the first of July 1878).

The conviction of Caverni has been completely ignored by all the historians who dealt with the diffusion of the Darwin's theories since the documents of the Holy Office and of the Congregation of the Index, regarding the debate on evolutionism, are available only since 1998 (date of opening the archives to scholars). This assertion must not be misinterpreted. In the meager biographies of Caverni the decree of the Congregation of the Index is mentioned but, we can say, only incidentally because, actually, all these biographies have been written with the intention of speaking about the author of the *History of the experimental method in Italy.* Today, due to a study of Artigas, Glick and Martinez *Negotiating Darwin* (2006), we know that Caverni, at that time, was the only victim of the Holy Office among the supporters of Darwin, instead.

Not even the work of Darwin itself was placed on the Index. In the strict sense, no actions of the Holy Office took place against evolutionism either, what many times maintained by the «Civilta Cattolica» notwithstanding. In the case of Caverni, the Holy See carefully avoided an official action against evolutionism for fear of recreating a Galilei case and then limited itself to insert Caverni's book in the *Index Librorum Prohibitorum* without publicly making the reasons explicit. Only father Giovanni Giovannozzi (1860-1928), about whom we shall speak below, in a warm recollection of Caverni goes into details saying that the decision of the Holy Office had been caused by attacchi piuttosto caustici ed acri [… ] ad Istituti, metodi e persone del mondo ecclesiastico [7] that Caverni launched in his book. Our author was not a man who skimped criticism when he thought it due but, to second-guess, we share the conclusions of the authors quoted above.

---

[6] ...... being run into this heated debate, which from a few years stirs people on the question of the origin of animal species, I wanted to stop for examining more calmly these new doctrines, fervently championed by an army of naturalists commanded by a celebrity, Charles Darwin. And checking his doctrines with the Genesis [......] I seem to have found that the new doctrines of the English naturalist, instead of being opposite to what can be read in the Book of Moses, fit into it in a more orderly way than the interpretations of the ancient exegetes do.

[7] the rather acid and caustic attacks [.....] against Institutions, methods and persons of the ecclesiastical world.

We are not allowed to know which was the impact on him of the condemnation to Index (we don't know if he has left some comments or other in his unpublished diaries). Father Giovannozzi wrote that he, come suo dovere, si sottomise pienamente[8].

Certainly, the thing did not provoke consequences nor controversies, not even in the lay field. On the other hand, at that time having a work placed on the Index did not imply serious difficulties to continue the sacerdotal office. The example of Antonio Rosmini was still alive, a priest philosopher having two works on the Index, but esteemed by the Popes who followed one another during his life.

Incidentally we note that Caverni was close to Rosmini as regards the general lines of his philosophical creed. It should also be said that, in the meanwhile, Caverni had begun to concentrate his studies on the history of science, in particular of mechanics and its applications. Already on April '75 he had received a letter of appreciation from the young man Antonio Favaro (1847-1922), at that time professor of graphic statics at the university of Padua and not even thirty years old, for the aforementioned book on the work of Galileo. This letter initiated a relation between Raffaello Caverni and Antonio Favaro destined to last for a period of fifteen years, which, undoubtedly, would have been the period of great intellectual fervor and scientific production of Caverni.

THE RELATIONSHIP WITH ANTONIO FAVARO AND THE TOMASONI AWARD

In the letter we mentioned above, besides to congratulate for the bellissimo lavoro[9], Favaro also asked if Galileo and his pupils had deepened the problem of the causes of earthquakes about which it was spoken in the book. In this way a dialogue begins between the two scholars who would have exchanged information on Galileo's manuscripts and the relative interpretation for years. Later on, Favaro introduced Caverni to the Roman prince Baldassarre Boncompagni, founder and publisher of the «Bullettino di Bibliografia e di Storia delle Scienze matematiche e fisiche», first work of this type in the world which was published for twenty years from 1868 to 1887. On the «Bullettino», Caverni will publish on September '78 a long essay (more than fifty pages) with the title *Notizie storiche intorno all'invenzione del termometro*[10]. This paper was his first work not being of popularization. In 1882, Antonio Favaro, who by now had become a well known scholar of Galileo also out of Italy, publishes his book *Galileo Galilei e lo Studio di Padova*[11] and asks Caverni for reviewing it.

The Author contented his friend (by this time their relation was a deep-rooted friendship) and published on the «Rassegna Nazionale» (n. 12, 1883) the asked review. But the friendship did not prevent Caverni from exposing his criticism. In fact, he observes in that review .... abbia il Favaro proseguito con più diligente studio quella prima parte, la quale riguarda la vita esteriore, che non l'altra spettante alia vita intima del pensiero, e che mentre in quella prima discute sempre pensatamente, e conclude con libertà di giudizio; in questa invece se ne stia molte volte contento a espor le cose, riposando sull'autorità di qualche altro scrittore[12].

---

[8] doing his duty, did give in completely.

[9] finest work

[10] *Historical Notes about the invention of the thermometer.*

[11] *Galileo Galilei and the Study of Padua*

[12] .... Favaro has taken greater pains in that first part, which regards the outward life, than the other one regarding the inward life of the thought, and whereas in the first one he always discusses thinkingly and

As an example, he then mentions the question of the isochronism of the oscillations of the pendulum and concludes: Se si vuole insomma perfidiare a dire che l'isocronismo del pendolo sia stato scoperto da Galileo, per via sperimentale, non ci è modo a scusarlo dalla taccia di osservatore o sbadato o poco sincero. Ma il fatto è che non fu l'esperienza occasione della scoperta, sì un corollario di geometria meccanica, e la tradizione della lampada oscillante nel Duomo di Pisa, io per me la credo una favola[13].

We have expatiated on this quotation with the aim of showing that, in not questionable times, Caverni could be already considered a *detractor* of Galileo, according to the criterion of his critics to come, although a green detractor. Favaro responded that, even if *the question of the lamp* is a legend, yet la leggenda è cosi bella, cosi seducente[14] that he did not feel up *to* spogliare la nostra tradizione scientifica della parte leggendaria[15]. But at this time Favaro had already initiated his attempts of giving rise to a great *literary enterprise,* in which he wanted to take Caverni into partnership. That was a new edition of Galileo's works.

Then, we see the beginning of the more complicate and at the same time more delicate period of the relationship Caverni - Favaro. This period has been studied and analyzed in detail by Cesare S. Maffioli, who has also published a regest of their correspondence (1985).

In 1882, Favaro initiated negotiations with the *Successori Le Monnier* for the edition he had in mind, according to the project he had already anticipated in appendix to his book *G. G. e lo Studio di Padova.* The negotiations would have lasted a couple of years and then everything came to nothing. In the meantime, Favaro published some manuscripts of Galileo (until then left unpublished and therefore not included in the Albèri edition) pertinent to the Pisan period of Galileo's studies of mechanics.

The two friends discussed for a long time about the writings of mechanics of the Pisan period, a part of which had been published by Galileo himself in his last work *(Discourses and Mathematical Demonstrations .... - 1638)* and another part was left unpublished together with some writings of the Paduan period. The discussion was about dating the unpublished manuscripts, some of which were fragments, and then to their insertion in the volumes to be published. Favaro also asked E. Wohlwill, at that time the most estimated German scholar of Galileo, for his opinion. While these scientific discussions advanced well, a second stage of negotiations with the reconstituted publishing house *Successori Le Monnier* went in a new failure, in spite of the intervention of the Ministry of Education. Later on, Favaro came to an arrangement with the Ministry which will finance the edition. The twenty volumes of what became the National Edition of Galileo Galilei's works were published by the publishing house *Barbera* in a limited number of copies (500). The volumes came out between 1890 and 1909.

But now we shall go back in our story. During the negotiations of Favaro for the edition of the works, little by little the relationship between the two friends became complicate. Caverni, who in these years was writing a book on *Dante' s Physics,* for whose edition he was denied a contribution by the Ministry, realizes that he was being progressively excluded

---

concludes with liberty of thought; in the second one limits instead himself to display the matter, following the authority of some other author.

[13] If one wants in a word to be perfidious and to say that the isochronism of the pendulum has been discovered by Galileo, by way of experiment, there is no way of saving him from the charge of careless or little sincere observer. But the fact is that the occasion of the discovery was not due to the experience, rather to a corollary of mechanic geometry and the tradition of the lamp oscillating in the cathedral of Pisa, in my opinion is a fable.

[14] the legend is so beautiful, so seductive
[15] despoil our scientific tradition of the legendary part.

from a protagonist role in the editorial staff of the National Edition. Favaro felt embarrassed to justify this exclusion and partly laid this responsibility on the Ministry.

At the same time, as if he wanted indemnify Caverni for the experienced disappointment, exhorted him to participate with his own writing in an award expiring on 31 March 1889 announced in second edition by the Royal Venetian Institute of Sciences, Letters and Arts. The prize, not awarded in the first edition because of the inadequacy of the competitors, consisted of lire 5.000 of that time due to a testamentary legacy of December 1879 to the Venetian Institute on behalf of such G. Tomasoni, to be destined to chi detterà meglio la storia del metodo sperimentale in Italia[16]. Caverni, once convinced to participate, produced a work whose manuscript, as attested by the Board of Examiners, di proporzioni veramente colossali (sono 3264 pagine di grandissimo formato tutte scritte per intero).[17] The Board consisted of A. Messedaglia (later on replaced by G. Lorenzini), A. Minich and Antonio Favaro (who wrote the final report). On December 1889, at request of Favaro, Caverni sent to the Board also a summary of his manuscript. Later on, this fact gave rise to a retrospective polemics. Strictly speaking, since the manuscripts to be examined had to be anonymous, in his capacity of a member of the Board Favaro should not have requested that summary, justifying his request by the excessive length of the manuscript, which made the exam of the Board difficult. The Council of the Royal Venetian Institute charged of the exam of the manuscripts presented at the Tomasoni Award, i.e. the above Board, awarded the prize to Raffaello Caverni on February 16th, 1890. At this point the Author began to worry, since the Institute had the rule of paying the prize amount only after the publication of the winning work and only after having received 50 presentation copies of it. Obviously such a rule had been fixed having in mind essays of the usual size, not works in several volumes as that of Caverni once printed was. We spare the reader the troubles of Caverni for succeeding in reconciling those rules with the publication of the *History*. Fortunately, he found a publisher Maecenas, Mr. Civelli, and the work began to come out starting from 1891, although in the midst of polemics. Just in the report of the Board, Favaro began the address which he will carry on for the subsequent decades, even if here the epithet of detractor has not yet come out.

Let us see L'Autore si manifesta senza reticenze ammiratore profondo di Galileo (e chi mai non lo sarebbe?); ma egli, forse posto in sull'avviso dall'ingiusto giudizio di chi volle esaltare Galileo con pregiudizio di tutti i contemporanei, e non consentendo in esso, pare quasi sempre in guardia contro conchiusioni che al sommo filosofo riescano soverchiamente favorevoli, ed il *rationabile obsequium,* che lo storico deve prefiggersi come massima indeclinabile, è da lui spinto, ci sia lecito il dirlo, ad un eccesso che noi reputiamo ingiustificato[18].

CAVERNI DETRACTOR

As we have already said, he succeeded to publish his work by help of senator Civelli, at that time owner also of a chain of newspapers, besides of a printing house: the first volume came out in 1891. He also had a highly respectable reviewer, no less than Giovanni Virginio

---

[16] whom will better write the history of the experimental method in Italy

[17] was of a truly colossal size (3264 pages of greatest format all wholly written)

[18] The Author presents himself without reticence as a profound admirer of Galileo (and who on earth should not be?); but he, perhaps forewarned by the unfair judgment of those who wanted to exalt Galileo to prejudice of all his contemporaries, and not agreeing, seems to be almost always on guard against conclusions which result excessively favorable to the supreme philosopher, and the *rationabile obsequium,* that a historian must propose to himself as an intransgressible rule, is driven by him, allow us to say, to an excess we deem unjustified.

Schiaparelli, the greatest Italian astronomer of that time, who was also an historian of ancient astronomy. The review of Schiaparelli, as a matter of fact an essay of almost thirty pages, was very favourable; although criticizing Caverni for a preconceived hostility towards Galileo and for some arbitrary judgments, he put in evidence the merits and the importance of the work, considering it certainly of high level. In fact, he defined the work il più gran corpo di storia scientifica che vanti la letteratura italiana[19]. Finally, let us come now to speak about Caverni's judgments on Galileo. Recall again what the cultural atmosphere and the myth of Galileo were at that time. Until then, the historians behaved like the doxographs to the philosophers and astronomers of the ancient Greece, mixing reality and legend in not always verifiable proportion. Thus, they attributed to Galileo the most disparate "inventions" and "discoveries", inducing to consider them as very important things, on which the greatness of Galileo depended.

Obviously, this gave rise to an image of Galileo not corresponding to the truth in the popular literature. Indeed, Caverni was the first one to be interested in the thought of Galileo and this led him to deny several groundless attributions. It also happened to him to go too far in doing this, sometimes, so to say, throwing out the baby with the bath water.

On the other hand, Galileo himself sometimes had claimed the priority of certain results and had not hesitated to start even harsh discussions with his opponents in the scientific and technical field. This negative side of Galileo's character has certainly not been well accepted by Caverni who, from himself, has emphasized his faults. We are not convinced of the explanation, substantially guaranteed also by Favaro, that attributes to Caverni a retaliation (on Galileo in his *History)* for having been excluded from the National Edition of the Works. We have already recalled his engagement in studying the chronology of the Galilean manuscripts concerning the mechanics. It must be also remarked that the problem was of noteworthy importance for the understanding of the genesis of certain concepts and this can be corroborated by the fact that it is considered still such (Stillman Drake, one of the most qualified scholars of the work of Galileo, devoted to this problem a long essay in 1979). We can undoubtedly say that, in the ideas on Galileo's mechanics, Caverni was a step further than his contemporaries, as we shall see later on. Unfortunately, the Author unexpectedly died on January 30[th], 1900, without having the possibility of finishing his work. The fifth volume had come out on 1898 and the sixth one will come to light later on, unfinished. As far as it is known, Favaro did not authored any review of published Caverni's work while Caverni was alive. In a certain sense, he began to do it in the obituary he read at the assembly of 25 February 1900 of the Royal Venetian Institute. Favaro recalled that, already in the report of the board, ... non si passavano tuttavia sotto silenzio alcune mende, dovute in parte al difetto di cognizione delle fonti straniere, ma soprattutto a certi preconcetti sulla interpretazione dei documenti; la quale non si stimò sempre scrupolosamente conforme alla sana critica e al rigore storico, per modo che egli fosse, fra le altre, condotto a raffigurarsi un Galileo non vero, né come uomo, né come scienziato[20].

Then he went on saying that Caverni had not paid attention to the suggestions of the board of examiners, on the contrary he had *strengthened the dose,* and the more, the published text was different from the manuscript in several parts. All that, and other, aforesaid, he went on saying ... ma la tomba che si e anzi tempo dischiusa per lui ha cancellato dalla mia memoria il triste ricordo delle ingiustificate recriminazioni e dei poco benevoli giudizi, non lasciandovi altro che il rimpianto profondo e sincero del suo grandissimo sapere e delle doti altissime della sua mente.
Ed invero, qualunque siano le critiche e le censure che potranno muoversi all'opera monumentale di Raffaello Caverni, essa restera pur sempre la piu ricca raccolta di materiali per la storia della Scuola Galileiana, la quale da nessun altro prima di lui era stata tanto ampiamente e dottamente illustrata[21].

---

[19] the greatest corpus of scientific history that the Italian literature boats

[20] ……. however, one did not pass over in silence a few flaws, partly due to a lack of knowledge of the foreign sources,but above all to certain preconceived ideas on the interpretation of the documents; which was not always scrupulously conforming to a sound criticism and to historical exactness, so that, among other things, he was led to image a Galileo not true, neither as a man, nor as a scientist.

This magnanimous *parce sepulto* did not last. Favaro had started in 1894 to write a series of memoirs concerning the *Friends and correspondents of Galileo* which would have ended with the forty-first memoir in 1919.

Already in 1904, in the memoir regarding Cesare Marsili, he began the series of quotations from the *History* of Caverni (which would have been about twenty in all in the years). Almost all these quotations would have turned into invectives.

It seems that Favaro had planned the mission of contesting Caverni's judgments violently on any occasion. This mission lasted until his death. In the above quoted memoir, Caverni is un tale che s'era proposto il triste compito di provarsi a sfrondare l'alloro immortale che cinge la fronte di Galileo, e a togliere fede alia unanime testimonianza essere stato il cuore di lui l'altezza somma della mente[22]. In a subsequent memoir (1912) concerning Viviani, he said:
... ha dato al maggior detrattore di Galileo una di quelle cosi avidamente cercate occasioni per dipingerlo al mondo come il più spregevole dei plagiari a danno dei suoi stessi discepoli, … [23]. Obviously, Caverni also has a front seat in an article on «La Rassegna Nazionale» (on February 16th, 1907) bearing the title *Ancient and modern detractors of Galileo,* where one starts from the contemporaries of Galileo, Italians and foreigners, and arrives in the last section to an italiano che sembra essersi assunto il triste compito di sfrondare a tutta possa l'alloro che cinge la fronte immortale dell'instauratore del metodo sperimentale, ed in alcuni ponderosi volumi, nei quali si fece a tesserne la storia non v'è bassa ingiuria, velenosa insinuazione, ch'egli abbia risparmiato a danno del morto per far dispetto ai vivi[24]. As one can see, Favaro was obsessed with the laurel. This botanical metaphor is recalled many times by him when speaking of Galileo who, usually, was mentioned as the *supreme philosopher* or yet as the *divine philosopher.* If we consider that, in the last decade of the XIX century and the next ones of the XX, Favaro was considered a person of unquestionable authority on any matter regarding Galileo, and this also justly by force of the forty years research he devoted to the *divine philosopher* and to the realization of the National Edition, then it is clear that his anathemas against Caverni should have consequence. In fact, Favaro has been followed by eminent components of the cultural establishment of that time. Roberto

---

[21] .... but the grave disclosed for him before time has sponged out of my memory the sad remembrance of the unjustified recriminations and of the not very benevolent judgments, leaving only the profound and sincere regret for his greatest learning and the highest endowments of his mind.
And really, whatever may be the criticism and the censures that one can put forward against the monumental work of Raffaello Caverni, it will still remain the richest collection of materials for the history of the Galilean School, which nobody before him had illustrated so widely and learnedly.

[22] someone who purposed the sad duty of trying to strip of leaves the immortal laurel which crowns the brow of Galileo, and of not believing any more the unanimous testimony which maintains that his hearth had been at the supreme highness of his mind.

[23] ...it has given to the greatest detractor of Galileo one of the so avidly sought chances of picturing him as the most despicable plagiarist in the word, injuring his pupils themselves....

[24] Italian who seems to have accepted the sad duty of stripping of leaves, with all his strength, the laurel which crowns the immortal front of the establisher of the experimental method, and, in some ponderous volumes where he was putting together all the story, he does not spare any rude insult and nasty insinuation to the prejudice of the dead in spite of the living ones.

Marcolongo, who yet had the merit of having brought to the attention of Duhem Caverni's work for what concerns the studies on Leonardo, did not refrain from mentioning Caverni many times as a *detractor* of Galileo. Aldo Mieli, aforementioned as the secretary of the Division of the History of the Sciences of the Society for the Progress of the Sciences, even if in the first volume of the review «Archivio di Storia della Scienza» (1920) of which he was the director, in a foreword to articles about Caverni's History, had said that it has a real and effective importance, later on (1937) mentioned Caverni as one of the declared enemies of Galileo in an international conference dedicated to the 3$^{rd}$ centenary of the work *Discourses Concerning two New Sciences*. The same label was obviously reserved to Duhem as well. And, finally, in the *Summary of the History of Scientific Thought* (1937) of F. Enriques and G. Santillana - in the Bibliography - the work of Caverni is quoted as a work whose biased judgments must be accepted with caution, in good company with Duhem. Obviously the list could continue, but we want to conclude with another writing of Favaro bearing the title *Apocryphal Galileo's writings* (1917), in a review founded by Gino Loria as an ideal continuation of the «Bullettino» of Boncompagni. In 1917 (this date is not completely certain since the circumstances have not been well clarified) the sixth volume of the *History* of Caverni came out printed in unfinished form, since it was interrupted half-way of a phrase at p. 464, and dated 1900. In reviewing, so to say, this volume, Favaro accused Caverni of having ascribed to Galileo things written by himself imitating the style of the Galilean dialogues and, more, remarked la insistenza ed il peggioramento nell'insano proposito di denigrare ad ogni costo Galileo fino al punto da rappresentarlo campione del peripatetismo in confronto degli stessi suoi oppositori peripatetici.[25]

The tone used by Favaro is particularly acrimonious, almost as if this volume had been printed to be rude to him. Caverni, even if dead, continued not to agree with him completely. Now, to round off the story, let's go on to see what happened after the famous agenda of the Division of History of the Sciences of April 17$^{th}$, 1919 which we have mentioned at the beginning. In the first volume we have already mentioned, Aldo Mieli's review published an article of father Giovanni Giovannozzi, Piarist and astronomer, former director of the Ximenian Observatory and member of the Italian Society of Physics, in which Caverni was affectionately defended, through admitting an unexplainable disposition of him a creder vere tante e sì gravi accuse contro la probità professionale e personale di Galileo[26].

Furthermore, an article of Carlo Del Lungo about the *History*, in which he tried to refute the work both on methodological and stylistic level (in point of fact the style was a little obsolete) ended exhorting to write monographs on Caverni's work.

In the same volume, there was an article of Favaro on the phases of Venus, a somewhat complex question of which Caverni gave the credit to Castelli, instead of Galileo. The second volume of the same review contained a new article of Carlo Del Lungo, this time on the pendulum and clock, another *glory* to be credited to Galileo. By then, the objective was attained. In Italy the work of Caverni would not have been considered as fundamental in the studies about Galileo and the history of mechanics by the scholars any more. The *Summary* of Enriques and de Santillana really summarizes the finally achieved public opinion. But things did not end that way.

---

[25] the insistence and the worsening on the insane intent of denigrating at any cost Galileo to such an extent of representing him as a champion of the Peripateticism compared to his Peripatetic opponents themselves.
[26] to believe so many and so heavy accusations against the professional and personal probity of Galileo

CAVERNI CLEARED

This verb seems to be particularly suitable since, after a period of forty years of segregation, the work of Caverni was actually cleared. In this regard one usually quotes a phrase of E. Garin, in his book *Science and Civil Life in the Italian Renassance,* (1965): a work injustly forgotten. But already in 1958 (in the *Boringhieri* edition of the Galileo's *Discourses,* edited by A. Carugo and L. Geymonat) A. Carugo, author of the notes, often analysed Caverni's interpretations considering him a sound interlocutor. In other words, even if too late, finally Caverni came again into play in the field of the scholars of Galileo. In the years between the two world wars, out of Italy, Caverni was held in repute by Koyré; for the scholars of English language the discovery will occur later on instead, like in Italy: we limit ourselves to mention Winifred L. Wisan and Stillman Drake among the main scholars. Since the seventies the list of the Italian scholars who came into contact with the work of Caverni became rather goodly, starting with Giorgio Tabarroni who devoted to him a biographic essay on «Physis» and edited a facsimile reprint of the work of Caverni for the publisher *Forni* of Bologna. Two years later, a new facsimile reprint would have been published by *Johnson Reprint* - New York. Now Caverni is definitively cleared. The Italian scholars of the last generation, and actually also of the second last one, do not ignore him and do not consider him any longer a *detractor,* but perhaps an *eccentric* they can still discuss with. We do not want to run the risk of forgetting somebody and then we shall not make a list of the works of the Italian scholars of the last decades. We limit ourselves to mention only two texts: the splendid volume *Galileo - La sensata esperienza* - (1988) edited by Paolo Galluzzi, with the contribution of Gianni Micheli and *Galileo Galilei* (2004) by Michele Camerota, where Caverni is quoted and discussed many times. At the end we recall the contribution, in English, of Giuseppe Castagnetti and Michele Camerota: *Raffaello Caverni and his History of the Experimental Method in Italy* in *"Galileo in context"* (ed. Jürgen Renn-2001).
Finally, what moral can we draw from this story?
Perhaps more than one, but all very obvious; it's a pity not having minded before! The fact that a trial, truly obscurantist, has been started by an assembly of scientists who were fighting for the progress of the science, makes us think about the recurrence, in some unexpected cases, of *the sleep of reason.*
The result, from a scientific point of view (that is, regarding the studies of history of science), is that Caverni's work has not been used in due time and in the due way and now it results however out of date, without being a classic.

BIBLIOGRAPHIC REFERENCES

RIASSUNTO

Raffaello Caverni, un prete cattolico, fu un vero intellettuale laico e anti-establishment nelle sue opinioni sia riguardo a Galileo che a Darwin.
Egli si oppose alla miticizzazione di Galileo, in atto nell'Italia post-unitaria, sebbene lo considerasse un grande scienziato.

Come conseguenza, la comunità scientifica dell'epoca, sotto l'influenza di Antonio Favaro, condannò aspramente la sua opera *Storia del metodo Sperimentale in Italia.*
In forza di ciò, l'opera del Caverni fu emarginata dal dibattito scientifico in Italia per almeno quarant'anni.

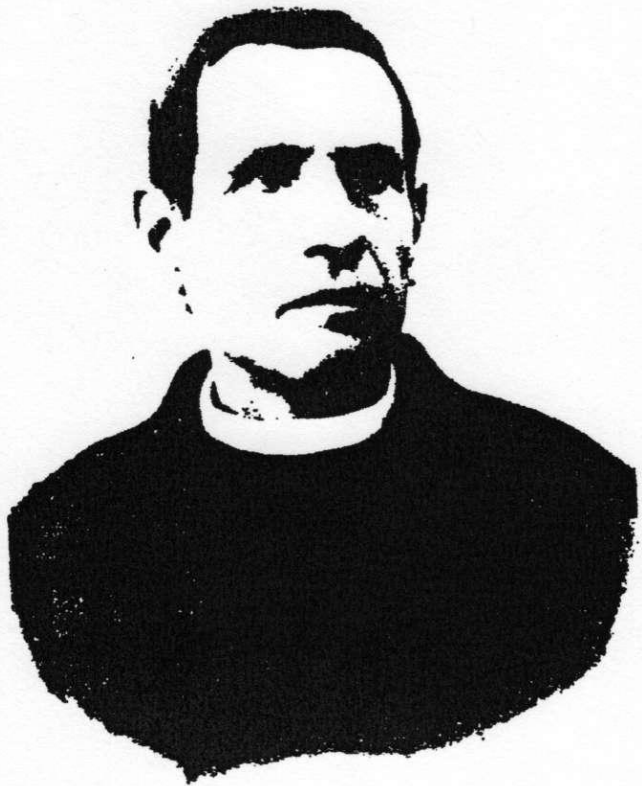

Fig. 1 – Raffaello Caverni – from the picture inserted in the paper of Aldo Mieli quoted in the text (Archivio di Storia della Scienza – vol. 1 p. 264 – 1920)

# STORIA

DEL

# METODO SPERIMENTALE

## IN ITALIA

OPERA
DI
RAFFAELLO CAVERNI

TOMO I.°

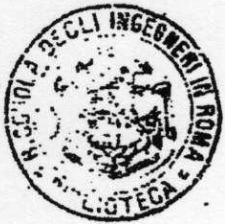 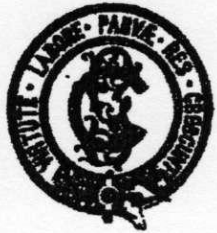

*FIRENZE*
STABILIMENTO G. CIVELLI - EDITORE

1891.

Fig. 2 - Title page of the first volume of the **History** ( from the copy of the original edition owned by the Library of the Mathematics Department of La Sapienza – University of Rome)

# STORIA
### DEL
# METODO SPERIMENTALE
## IN ITALIA

OPERA
DI

RAFFAELLO CAVERNI

TOMO VI.°

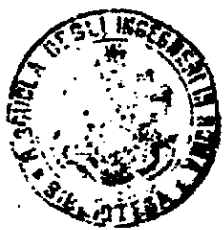 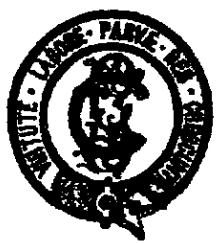

*FIRENZE*
STABILIMENTO G. CIVELLI - EDITORE

1900.

Fig. 3 - Title page of the sixth volume of the **History** appeared, maybe, on 1917, but dated 1900

464 *Storia del metodo sperimentale in Italia*

terato, com' egli è, nella forma che l' ha stampato il Manolessi » (MSS. Cim., T. XXIII, fol. 22):

Il qual Manolessi, ricevutone così il comando, sospese la pubblicazione, e spedì la copia desiderata, avuta la quale in mano è naturale che il Principe ricorresse con l'occhio e col pensiero alla proposizione seconda del secondo libro, e al trovarla stampata conforme al manoscritto si deve essere risovvenuto del Torricelli, e come gli avesse, 17 anni fa, fatto osservare che se la nuova acqua nel regolatore del fiume sta in altezza alla prima come quattro a due; non però come quattro a due staranno le velocità respettive, ma come quattro alla radice di due. Dev' essere inoltre esso Principe stato informato come, risaputa l' osservazione, il Castelli rispondesse, che sebben non si trovasse sodisfatto della dimostrazione, nonostante la proposizione in sé stessa, essendo il legittimo resultato dell' esperienza, non poteva non esser vera. Ond' essendo dovuto convenir di ciò il Torricelli, non rimaneva dubbio intorno alla parte della detta proposizione, che aveva bisogno d' esser corretta, secondo le convenzioni stesse fatte fra que' due grandi uomini. La difficoltà però consisteva nel saper trovare la ragion di un fatto particolare, che si sottrae alle leggi universali de' corpi naturalmente cadenti, per cui, essendo in quel punto presente in Firenze il Borelli, volle il principe Leopoldo conferir la cosa primieramente con lui, comandandogli di dirne il suo parere. Il Borelli allora rispose che questo sarebbe di sopprimere la dimostrazione, e in carattere corsivo stamparvi invece un avvertimento, che dicesse come quella mancava, perchè l' Autore fu sorpreso dalla morte, mentr' era in cercarla, e che perciò aveva pensato di supplirvi uno scolare di lui, mettendola in fondo al libro. Non decidendo il Principe nulla ancora del resto, comandò al Borelli facesse egli stesso quella dimostrazione, che pochi giorni dopo recapitava in palazzo, scritta in questa maniera:

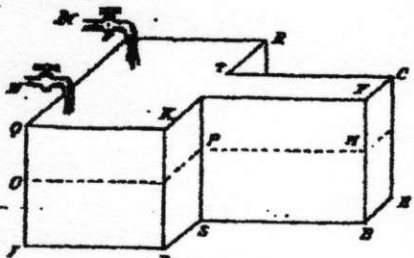

Figura 232.

« Sia il fiume SBC (fig. 232) per il regolatore CEBF, annesso al vaso QIDR, che sia prisma con le sponde erette all'orizzonte. E prima, l' origine del fiume M versi tanta acqua, che arrivi al livello OP, e scorrendo con la velocità S faccia nel regolatore la sezione rettangolare EBH. Poi l' altro sifone o torrente N, versi nuova acqua, ed arrivi al livello QR, e scorrendo con la velocità T per il fiume riempia la sezione rettangolare EF. Dico che la velocità T, alla velocità S, ha l' istessa proporzione che l' altezza FB, all' altezza HB. »

« La quantità d' acqua, che passa per la sezione EF, cioè il prisma acqueo QIDR, alla quantità dell' acqua, che passa per la sezione EH, cioè il prisma acqueo OIDP, ha l' istessa proporzione che l' altezza QI all' altezza OI, per avere i detti prismi la base comune. Di più, la velocità T, con la

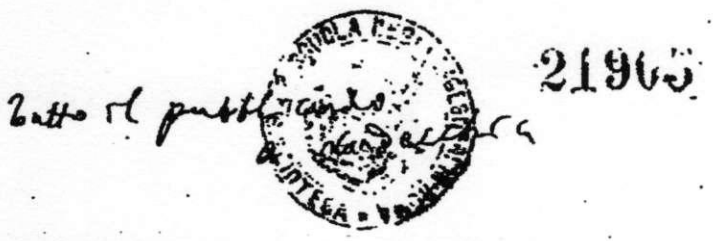

21965

Fig. 4 - Last page of the sixth volume of the **History**. In the curious sentence at the bottom, the bookseller A. Nardecchia guarantees that the book consists of all the published material

18